\documentclass[aps,pre,showpacs,twocolumn,showkeys,amssymb,floatfix,superscriptaddress]{revtex4}
\usepackage{graphicx}
\usepackage{epsf}

\begin{document}

\title{The dynamics of opinion in hierarchical organizations}

\author{M. F. Laguna}
\email{lagunaf@cab.cnea.gov.ar}

\affiliation{Centro At\'{o}mico Bariloche, CONICET and Instituto
Balseiro, 8400 San Carlos de Bariloche, R\'{\i}o Negro, Argentina}

\author{S. Risau Gusman}
\email{srisau@if.ufrgs.br}

\affiliation{Instituto de F\'{\i}sica, UFRGS, Caixa Postal 15051,
91501-970 Porto Alegre, RS, Brazil}

\author{G. Abramson}
\email{abramson@cab.cnea.gov.ar}

\affiliation{Centro At\'{o}mico Bariloche, CONICET and Instituto
Balseiro, 8400 San Carlos de Bariloche, R\'{\i}o Negro, Argentina}

\affiliation{Consortium of the Americas for Interdisciplinary
Science, University of New Mexico, Albuquerque, New Mexico 87131,
USA}

\author{S. Gon\c{c}alves}
\email{sgonc@if.ufrgs.br}

\affiliation{Instituto de F\'{\i}sica, UFRGS, Caixa Postal 15051,
91501-970 Porto Alegre, RS, Brazil}

\affiliation{Consortium of the Americas for Interdisciplinary
Science, University of New Mexico, Albuquerque, New Mexico 87131,
USA}

\author{J. R. Iglesias}
\email{iglesias@if.ufrgs.br}

\affiliation{Instituto de F\'{\i}sica, UFRGS, Caixa Postal 15051,
91501-970 Porto Alegre, RS, Brazil}

\date{\today}

\begin{abstract}
We study the mutual influence of authority and persuasion in the
flow of opinion. Many social organizations are characterized by a
hierarchical structure where the propagation of opinion is
asymmetric. In the normal flow of opinion formation a high-rank
agent uses its authority (or its persuasion when necessary) to
impose its opinion on others. However, agents with no authority
may only use the force of its persuasion to propagate their
opinions. In this contribution we describe a simple model with no
social mobility, where each agent belongs to a class in the
hierarchy and has also a persuasion capability. The model is
studied numerically for a three levels case, and analytically
within a mean field approximation, with a very good agreement
between the two approaches. The stratum where the dominant opinion
arises from is strongly dependent on the percentage of agents in
each hierarchy level, and we obtain a phase diagram identifying
the relative frequency of prevailing opinions. We also find that
the time evolution of the conflicting opinions polarizes after a
short transient.
\end{abstract}



\maketitle

\section{Introduction}
\label{intro}

The exchange of opinions plays an essential role in decision
making at all levels of social activities, from domestic matters
within the family, through business at all scales, to political
issues affecting a nation. The dynamics of opinion formation and
transmission, as well as the generation of consensus has attracted
the attention of mathematicians and physicists. This is probably
due to the fact that the emergence of a collective state is a
feature reminiscent of the behavior of many-particle systems in
the fields of physics, chemistry and biology. Therefore, the
formation of public opinion has been the subject of a number of
recent works that, to some extent, capture the fundamental
processes that determine the emergence or not of consensus within
a population
\cite{weisbuch02,stone61,hegselmann02,laguna03,laguna04}. In Refs.
\cite{weisbuch02,stone61} the condition to reach consensus in well
mixed populations is analyzed. Additional results of similar
models have been shown in Refs. \cite{laguna03,laguna04}, such as
the relevance of the dynamical time scale on the formation of
minorities and the role of a complex topology in the contact
network. A more general class of models has been studied in Refs.
\cite{hegselmann02}, in systems of completely connected agents.
Though, there is a feature ---not analyzed in those
contributions--- that plays and important role in many social
systems, and which certainly affects the dynamics of the
propagation of opinion: hierarchy. A hierarchical organization
characterizes many social institutions, ranging from families to
companies or governments, armies or churches. The emergence of a
hierarchical structure in social networks is, by its own, the
subject of related studies, such as those found in
\cite{bonabeau95,stauffer03,boguna03,copelli02}. In a hierarchical
system, the interaction between two individuals is asymmetric if
one of them belongs to a higher level of the hierarchy.
Consequently, it can exploit its authority to impose its opinion
to the other one, regardless of the actual value of this opinion.
Nevertheless, the lower levels of the hierarchy may well be able
to impose \emph{their} opinion to the higher ones, if they are
persuasive enough. The situation is, to some extent, similar to
ecological systems involving competition between species that
display different colonization capabilities: the worse competitor
may thrive if it is a better colonizer \cite{tilman94}.

The purpose of the present paper is to propose a simple model for
the flow of opinions in a hierarchical social system. In this
model, the agents are organized in a hierarchy that determines
their degree of authority in any pair interaction. Besides, each
agent is endowed with a persuasion ability that may allow it to
convey its opinion to another agent, even against authority.

\section{Model}
\label{model}

We consider a population of $N$ interacting agents, each of them
characterized by its {\it opinion}, its {\it persuasion}
capability and belonging to an {\it authority} stratum. While the
opinion can be changed through interactions between agents, the
authority and the persuasion are assumed to be invariable. The
agents are distributed in a hierarchy of $r$ authority levels
enumerated from $1$ (lowest authority) to $r$ (highest authority).
We call $x_i$, with $1 \leq i \leq r$, the fraction of the
population that belongs to each level of the hierarchy. We assign
to each agent a value of persuasion, $p_i$, chosen at random
between $0$ and $1$.  Initially each agent has its own opinion
which is represented by its index, from $1$ to $N$, i.e.
$O_i(t=0)=i$.

Since we are interested in the study of a population in which the
hierarchy imposes rules of interaction, persuasion is not the only
way to convince an agent. Indeed, agents with very different
authority levels do not have a symmetric interaction. We model
this fact by allowing an agent to convince another one only if the
authority of the convincer is higher, equal to, or one authority
level lower than that of the partner.

At each time step we pick an agent ($i$) at random which will try
to convince a partner by imposing its value of opinion. The second
agent $j$ is selected at random from the same or a lower level or
the immediately upper one ($a_j \leq a_{i}+1$). The interaction
between agents is as follows: agent $i$ tries to convince agent
$j$ using first its persuasion, in such a way that the opinion of
$j$ will become that of $i$, i.e. $O_j(t+1)=O_i(t)$, with
probability $p_i$.  If persuasion fails, agent $i$ will try to
impose its authority: if $a_i > a_j$ then $O_j(t+1)=O_i(t)$;
otherwise, no change occurs. The final result of this process is a
consensus of opinions. This means that in the stationary state all
agents share the same opinion. The value of the resulting dominant
opinion identifies the agent that has convinced the whole
population, which we call the {\it leader}.

\subsection{Numerical simulations}
\label{numer}

\begin{figure}[tbp]
\centering
\resizebox{\columnwidth}{!}{\includegraphics{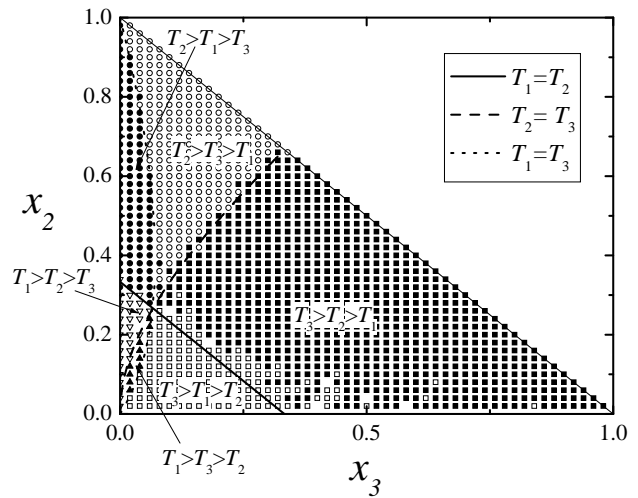}}
\caption{Phase diagram showing the relative frequencies of the
level from which the leader arises. Labels $x_2$ and $x_3$ indicates
the fraction of the population in the intermediate and upper
levels, respectively. The lines represent the results of the mean
field model with $\beta_{31}=1$ and $\beta_{13}=0$.} \label{fig1}
\end{figure}

We perform numerical simulations for an $r=3$ hierarchy, which is
the minimum number of authority levels that gives nontrivial
results and also allows us to build a two dimensional diagram. We
set the fraction of agents in each level of authority and assign
their values of persuasion at random. For all the initial
conditions and random number sequences studied, the simulations
always arrive to a final state where all the agents share the
dominant opinion. It is worth to emphasize that for a given
allocation of agents among the authority strata, the values of the
persuasion are attributed at random, therefore the leader can
belong to different authority levels for different initial
conditions. Then, for characterizing the leader's origin we
consider an ensemble of samples and perform an average over them.
In Fig.~\ref{fig1} we plot typical results for a system of $N=100$
agents and $1000$ samples. The horizontal axis indicates the
fraction of the total population that has the maximum authority,
$x_3$, whereas the vertical axis measures the fraction of agents
in the intermediate level of authority, $x_2$. The label $T_{i}$
represents the fraction of samples in which a member of the $i$-th
level becomes the leader. In region labeled with
$T_{i}>T_{j}>T_{k}$ the most frequent leader belongs to the $i$-th
level of authority, followed by leaders from the $j$-th level. One
can observe that in a large amount of this phase space (filled
with squares) the leader belongs to the highest authority group,
as it is expected. However, when the fraction of agents in the top
level is below $x_{3}=1/3$, the intermediate stratum may impose
one of their opinions (region indicated by circles). Finally, when
there are very few top authority agents and a fraction $x_{2}$ of
intermediate agents lower than $1/3$, the opinion of the lowest
stratum may overpower the top ones (region indicated by
triangles). The lines in the figure separate different regions
obtained by means of analytical results that we discuss in the
next subsection.

\begin{figure}[tbp]
\centering
\resizebox{\columnwidth}{!}{\includegraphics{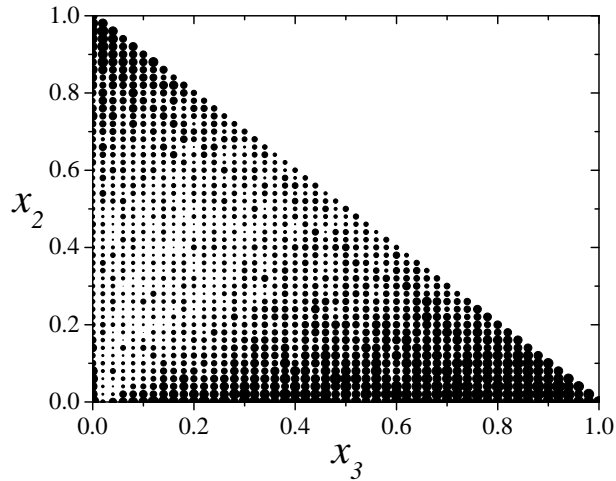}}
\caption{Map of the average persuasion of the leader. The axis
labels are the same as in Fig. \ref{fig1}. The smallest dots
correspond to an average persuasion $\langle p\rangle = 0.56$
whereas the largest ones correspond to $\langle p\rangle = 0.68$}
\label{fig2}
\end{figure}

Furthermore, to complete the characterization of the phase space,
we plot in Fig.~\ref{fig2} a map of the average persuasion of the
leader agent corresponding to the points of the diagram of
Fig.~\ref{fig1}. The smallest dots correspond to an average
persuasion $\langle p\rangle = 0.56$ whereas the largest ones
correspond to $\langle p\rangle = 0.68$. We remark that the
darkest regions in the diagram correspond to situations where most
of the population has the same value of authority. This diagram
reflects two facts: a) when the majority of the agents has the
same authority, one of the more persuasive becomes the leader
---even the agents with the highest authority need a persuasion
above the average to impose their opinion--- and b) in the region
where all levels have similar population, the authority prevails
over the persuasion, although the average value of the leader's
persuasion is bigger than the average persuasion over the
population ($\langle p\rangle = 0.5$).

\begin{figure}[tbp]
\centering
\resizebox{\columnwidth}{!}{\includegraphics{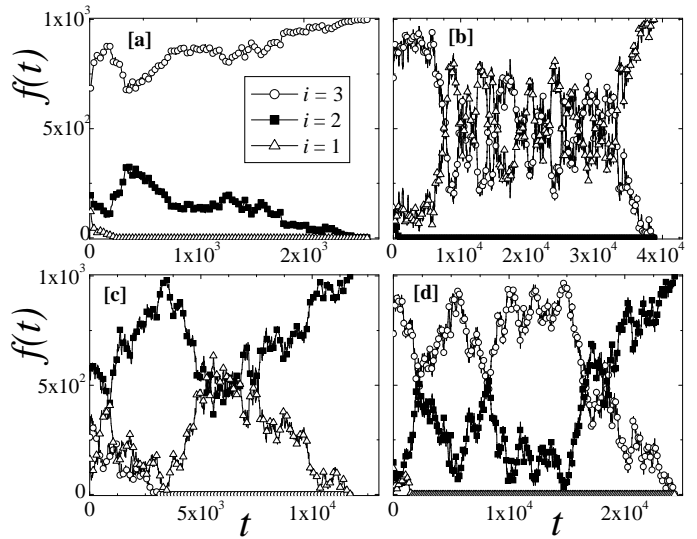}}
\caption{Time evolution of the number of agents sharing the
opinion of an agent belonging to a given authority level, for a
distribution in the hierarchy $x_{3}=3/5$, $x_{2}=1/5$ and
$x_{1}=1/5$. Circles indicate the number of agents that have an
opinion of an agent with authority $a_{i}=3$, squares correspond
to $a_{i}=2$ and triangles to $a_{i}=1$.} \label{fig3}
\end{figure}

Whereas the phase diagram shown on Fig.~\ref{fig1} tells us which
is the authority stratum from where the leader comes, the
examination of the time evolution that leads to this final state
may improve the comprehension of the mechanisms involved. Since we
know the authority level from which each opinion comes from, we
can follow the evolution of the number of agents that share an
opinion corresponding to each stratum. This temporal evolution is
shown in Fig.~\ref{fig3}, for a system of $N=1000$ agents. Each
curve corresponds to the number of agents sharing the opinion of
an agent belonging to the highest authority level (circles),
intermediate level (squares) and lowest level (triangles). Each
plot corresponds to a different sample with the same distribution
of population in the three levels, $x_{3}=3/5$, $x_{2}=1/5$ and
$x_{1}=1/5$. In these figures we do not distinguish between
different opinions within a given authority level. We observe
that, after a short transient, the system becomes polarized.
Agents belonging to one of the authority strata are convinced by
agents by one of the other two levels. So, opinions originated in
this stratum quickly disappear, while agents in the other two
levels keep struggling for a longer period (see, for example, in
Fig.~\ref{fig3}[b] the struggle between level $3$ vs. level $1$).
The final result depends on the sample: while in
Fig.~\ref{fig3}[a] an agent from level $3$ becomes the leader, in
Fig.~\ref{fig3}[d] the leader comes from level $2$. It is
remarkable that in Fig.~\ref{fig3}[c] the fight is between the two
lower levels, while the opinion of the highest one disappears
after a relatively long transient. This polarization seems to be a
verification of what happens in many election contests when the
public opinion converges to the two strongest candidates.

\subsection{Mean field analytical solution}
\label{mean}

A simple model for this system can be solved within a mean field
approximation. As we are mainly interested in the authority
stratum where the leader comes from, we consider that all the
agents in a given level have the same opinion, i.e., there are
only as many opinions as authority levels. This is equivalent to
first ``thermalize'' the opinions within each authority stratum
before to put them in interaction. This last process allows
individuals to pass from one opinion group to the other.

The mean field approach implies that at each time step we compute
the averages of the quantities of interest, which in turn depend
only on their averages in the preceding steps. We are interested
in obtaining the average number of agents in the different opinion
groups. It is clear that, although initially opinion groups and
authority groups coincide, as the system evolves each opinion
group will have members with different levels of authority. If we
call $G_{ij}$ ($1 \leq (i,j) \leq r$) the subgroup that contain
agents with authority $i$ and opinion $j$, the dynamics of the
system can be fully described by studying the evolution of
$g_{ij}(t)$, which is the population, at time $t$, of the
subgroups $G_{ij}$.

To perform the mean field calculation, we assume that two agents,
belonging to authority levels $k$ and $l$, are chosen at random.
The agent in $k$ convinces the agent in $l$ with probability
$\beta_{kl}$. That is, there will be a transition from the group
$G_{la}$ to the group $G_{lb}$ with probability $\beta_{kl}$,
where $a$ is the opinion originally held by the agent in $l$ and
$b$ is the opinion it adopted (that of the first agent). Notice
that the opinion of the first agent is not affected by the
dynamics.

The model introduced in Section~\ref{numer} corresponds to the
case where $\beta_{kl}=0$ for $|k-l|>1$. When the relation between
the authority levels does not rule out the interaction, one must
consider two cases: $k>l$ and $l-1 \leq k \leq l$. In the first
case, the agent in $l$ always changes its opinion because its
authority level is lower than $k$; thus, $\beta_{kl}=1$. For $l-1
\leq k \leq l$ we assume that the effect of persuasion is averaged
out in the population of each level, and therefore $\beta_{kl}$ is
simply the average of the persuasion of the agents in level $k$.
Thus, for a uniform distribution of persuasion (as considered in
the simulations above), we have that $\beta_{kl}=1/2$ for $l-1
\leq k \leq l$. The fact that the matrix $\beta_{ij}$ does not
change in time implies that, neglecting finite size effects, the
distribution of persuasion in each subgroup remains unaltered.
This assumption only holds if the initial distribution is the same
in each subgroup. In our model this is the case, because the
interaction does not depend on the persuasion of the agent who
(eventually) changes its opinion. Then, the probability that it
leaves the subgroup does not depend on his own persuasion.

Since the dynamics cannot change the total number of agents in
each level of authority, there are $r$ constraints: $\sum_{j=0}^r
g_{ij}=x_i$, for all $i$, where $x_i$ is the fraction of agents
with authority index $i$, as defined previously.

By considering, at time $t+1$, the probabilities that members of
groups other than $G_{ij}$ enter or leave this group, one can
write the evolution equation for all the subgroups. Let ${\bf O}_i
(t)=(g_{1i}(t),g_{2i}(t),\dots ,g_{ri}(t))$ be the vector of
populations of the opinion group $i$. Its time evolution is given
by (see Appendix):
\begin{equation}
\label{eq.evolution} {\bf O}_i(t+1)=(\mathbf{I}+\mathbf{A}/N) {\bf
O}_i(t)\,\,\, ; \, \,\, 0<i<r. \label{eq:mf}
\end{equation}
\noindent where $\mathbf{I}$ is the identity matrix, and the
matrix $\mathbf{A}$ is given by:
\begin{eqnarray}
\label{eq.A} A_{ii} &=& -\sum_{k \neq i} x_k \beta_{ki},
\\
A_{ij} &=& x_i \beta_{ji}, \,\,\, i \neq j.
\end{eqnarray}

Notice that the matrix $\mathbf{A}$ is not symmetric. Moreover, it
has only negative or null eigenvalues because the population of
each opinion group must remain bounded. Indeed, it has one null
eigenvalue, corresponding to the eigenvector ${\bf O}=(x_1,\dots
,x_r)$. This is evident from the fact that the number of agents in
each authority group must be conserved: $\sum_{i=0}^{r} {\bf O}_i
(t+1) = \sum_{i=0}^{r} {\bf O}_i (t)$. Except for very special
values of the parameters, this will be the only eigenvector
corresponding to the null eigenvalue~\cite{note}.

For large values of $N$ equation (\ref{eq.evolution}) can be
written as:
\begin{equation}
\label{eq.evt} {\bf O}_i(t)=\exp(\mathbf{A} t/N) {\bf O}_i(0)
\,\,\,;\,\,\, 0<i<r,
\end{equation}
where $O_{kl}(0)=\delta_{kl} x_l$. To obtain an expression
independent of the number of agents $N$ we rescale the time, $t
\rightarrow t N$.

As a consequence of the fact that $\mathbf{A}$ has only one
nonnegative eigenvalue, the calculation of the asymptotic
populations of each subgroups is an easy task: in the limit $t
\rightarrow \infty$, the evolution matrix
$(\mathbf{I}+\mathbf{A}/N)^t$ has only one non-vanishing
eigenvalue, whose value is 1. Thus, $\lim_{t \to \infty}
(\mathbf{I}+\mathbf{A}/N)^{t}=\mathbf{P}^{-1}\mathbf{Q}\mathbf{P}$,
where $\mathbf{Q}$ is a matrix whose only non-zero element is
$Q_{11}=1$, $\mathbf{P}^{-1}$ is the matrix of right eigenvectors
of $\mathbf{A}$ and $\mathbf{P}$ is the matrix of its left
eigenvectors. As we need only the eigenvectors corresponding to
the non-zero eigenvalue and, as it was mentioned above, its right
eigenvector is already known, we only have to calculate the left
eigenvector; i.e. $\bf v$ such that ${\bf v} \mathbf{A}=0$. This
eigenvector is very easily obtained by taking the vector product
of any $r-1$ columns of the matrix $\mathbf{A}$. With this
procedure, ${\bf v}$ is obtained up to a multiplicative constant
$C$ which can be calculated by imposing the restriction that the
sum of all populations is equal to $1$. Consequently, the
asymptotic populations are given by:
\begin{eqnarray}
\label{eq.asy} g_{ij}(\infty) &=& C \, x_i \, x_j \, v_j, \,\,
0<i,j<r, \\ C &=& \sum_{j=0}^{r} x_j \, v_j.
\end{eqnarray}

We have studied in detail the case $r=3$ and determined a phase
diagram by varying the fraction of agents in authority levels $2$
and $3$, as in Fig.~\ref{fig1}. Our first example is designed to
compare with the previous numerical results: while agents with the
highest level of authority can fully influence the lowest, the
opposite interaction is not allowed, i.e. $\beta_{31}=1$ and
$\beta_{13}=0$. The analytical results are plotted together with
the numerical ones on Fig.~\ref{fig1}. The final state of the
system, in the mean field approximation, is a coexistence of
opinions, where their relative weights are equivalent to the
frequencies of leader opinions in the simulation (averaging over a
large number of individual samples). Therefore, the label $T_i$
must be interpreted as representing the fraction of agents that
endorses the opinion $i$ in the asymptotic limit and, similarly,
$T_{i}>T_{j}>T_{k}$ means that, in this region, the prevailing
opinion corresponds to the $i$-th authority level, followed by the
opinions from $j$-th and $k$-th levels. It is interesting to
notice that, even in this extreme case, there are values of the
parameters for which the lowest authority opinion can become the
prevailing one. It is also remarkable the agreement between
analytical results (indicated by lines in the diagram) and
numerical ones (indicated by geometrical symbols). Indeed, almost
all the phase separations coincide, with the only exception of the
transition from $T_{3}>T_{2}>T_{1}$ to $T_{3}>T_{1}>T_{2}$. In
this case, the difference is restricted to the second placed in
the ranking of opinions where there is a significative shift in
the frontier line. This can be due either to a lack of statistics
in the simulations, given that the line is not well defined, or to
a limitation of the mean field approach.

\begin{figure}[tbp]
\centering
\resizebox{\columnwidth}{!}{\includegraphics{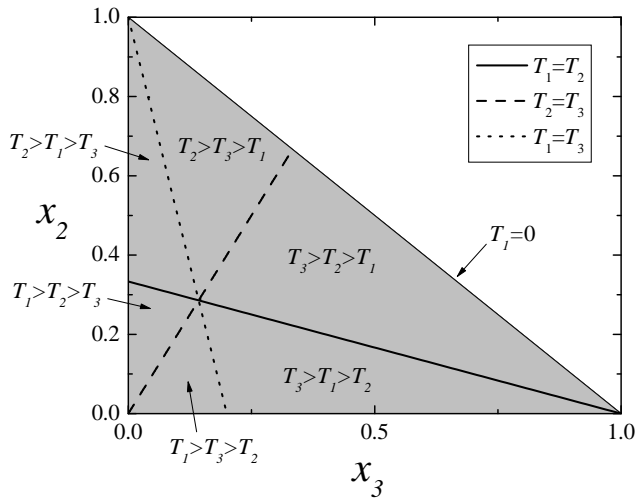}}
\caption{Phase diagram showing the mean field results of the
relative frequencies of the level from which the leader arises.
The transition probabilities are: $\beta_{13}=\beta_{31}=0$ while
$\beta_{ij}=1/2$ for all the other cases. Labels $x_2$ and $x_3$
indicates the fraction of the population in the intermediate and
upper levels, respectively.} \label{fig4}
\end{figure}

As a second example, we consider the case where the highest and
lowest levels of authority cannot communicate, i. e.,
$\beta_{13}=\beta_{31}=0$. To take into account the influence of
authority, we consider that an agent will convince another with a
lower authority, but the opposite will only happen with
probability $1/2$: $\beta_{32}=\beta_{21}=1$,
$\beta_{12}=\beta_{23}=1/2$. The results are shown in
Fig.~\ref{fig4}. Notice that if the intermediate level is not
populated no evolution is possible. When comparing the results of
Fig.~\ref{fig4} with those of the previous case we observe some
qualitative changes in the transitions lines $T_1=T_3$ and
$T_1=T_2$. Both changes happen in the region of small $x_2$, that
is small population of the intermediate level.

\begin{figure}[tbp]
\centering
\resizebox{\columnwidth}{!}{\includegraphics{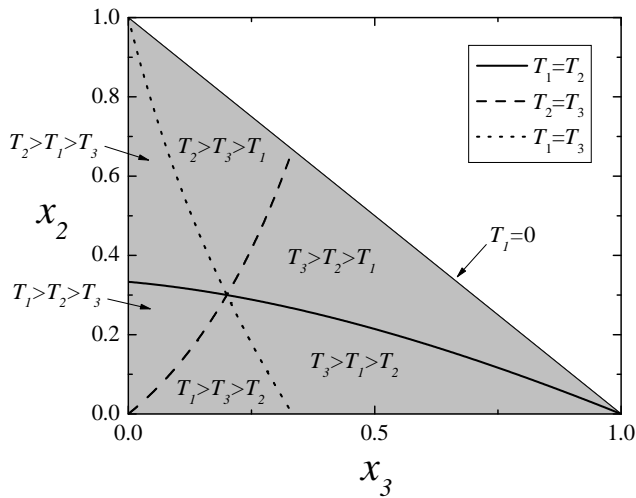}}
\caption{Phase diagram showing the mean field results of the
relative frequencies of the level from which the leader arises.
The transition probabilities are: $\beta_{13}=1/2$, $\beta_{31}=1$
while $\beta_{ij}=1/2$ for all the other cases. Labels $x_2$ and
$x_3$ indicates the fraction of the population in the intermediate
and upper levels, respectively.} \label{fig5}
\end{figure}

As a final example, we allow the highest authority to fully
influence the lowest ($\beta_{31}=1$) while the opposite happen in
only half of the interactions ($\beta_{13}=1/2$). The
corresponding phase diagram is shown in Fig.~\ref{fig5}, where we
verify that the resulting effect is an increase in the influence
region of the lowest level for low $x_2$ and an increase of the
influence of the highest authority level for high $x_2$.

\section{Conclusions}

We have described a simple model of opinion formation in a society
where both persuasion and authority are considered. The model has
the advantage of having an analytical mean field solution that
exhibit a fair coincidence with the numerical results. We have
concentrated in the three-authority level case as it is the one
with the minimum number of levels permitting to explore the
hypothesis of the model. Also, in this case, we are able to plot
the results in a two-dimensional phase diagram.

When the higher authority level can impose its opinion on all the
others (Fig.~\ref{fig1}) it is clear that its opinion prevails up
to the case when just a small fraction of the population belongs
to this level, providing that the population of the intermediate
level is also small. As the intermediate level acts as a
transmitter, when the fraction of occupation of this level
increases it is the opinion of the lowest one that becomes the
leader. Only when the percentage of occupation of  the
intermediate level is bigger than $30\%$ the opinion of the
intermediate level prevails. The results make explicit the
importance of the ``middle class'' as a transmitter to modify the
ruling opinions. Moreover, the diagram of Fig.~\ref{fig2} suggest
that even among the agents belonging to the highest authority
group, the persuasion determines who will impose its opinion.

The other cases studied (Figs~\ref{fig4} and \ref{fig5}) show that
either cutting the direct communication between the highest and
lowest levels, or by letting the lowest level to persuade the
upper one, the influence of this lowest stratum of the population
is highly boosted, generally at the expenses of the intermediate
one.

The model can be improved by considering the stubbornness
(resistance to change its opinion) of the agents, as well as the
social mobility that will permit a successful persuading agent to
go to higher authority strata. We are conducting studies in these
directions, as well as considering a higher number of levels in
the hierarchy.

\section{ACKNOWLEDGMENTS}

M. F. Laguna thanks the Solid State Group of Centro At\'{o}mico
Bariloche for the use of computational facilities and the
hospitality of the Instituto de F\'{\i}sica, Universidade Federal
do Rio Grande do Sul, Porto Alegre, Brazil. S. Risau-Gusman
acknowledges support from the Centro Latinoamericano de
F\'{i}sica. G. Abramson acknowledges support from Fundaci\'{o}n
Antorchas (Argentina). J.R. Iglesias acknowledges support from
Conselho Nacional de Desenvolvimento Cient\'{\i}fico e
Tecnol\'{o}gico (CNPq, Brazil). The authors thank D.H. Zanette for
valuable discussions and acknowledge partial support from CAPES
(Brazil) and SETCYP (Argentina) through the Argentine-Brazilian
Cooperation Agreement BR 18/00 as well as the hospitality of the
Consortium of the Americas for Interdisciplinary Science,
University of New Mexico during the final stages of the work.

\section{Appendix}

We assume that there are enough agents in each subgroup to have a
uniform distribution of internal variables. At each time step two
agents are chosen at random, and only the opinion of the second
agent will be changed, according to a rule depending on the
authority labels and on the internal variables. On average, this
implies that there is a probability $\beta_{ij}$ that the second
agent, of authority $j$, will change its opinion to that of the
first agent (of authority $i$).

In order to obtain the dynamics of the subgroup $G_{ij}$, we
calculate the probability that members of the different subgroups
enter $G_{ij}$ and the probability that agents from $G_{ij}$ leave
it for different subgroups. Transitions are only possible between
subgroups with the same authority index, as this is a fixed
variable. But these transitions can be induced from agents from
different authority groups.

Let us consider first the transitions induced by agents belonging
to the same authority group. The probability of an interaction
where one member of $G_{ij}$ convinces a member of $G_{ik}$
(transition $G_{ik} \rightarrow G_{ij}$) is $g_{ij} g_{ik}
\beta_{ii}$, and the result is the addition of one member to
$G_{ij}$. With the same probability the opposite interaction
takes place, i. e. one member of $G_{ik}$ convinces a member of
$G_{ij}$, depleting $G_{ij}$ by one. Thus, both fluxes
compensate, and they do not influence the dynamics.

When considering transitions induced from different authority
groups, the asymmetry becomes manifest. The probability that a
member of $G_{kj}$ induces the transition $G_{il} \rightarrow
G_{ij}$ is $g_{kj} g_{il} \beta_{ki}$, whereas the opposite
transition, as induced by an agent from $G_{kl}$ happens with
probability $g_{ij} g_{kl} \beta_{ki}$.

By taking into account all the interactions that add or remove one
agent to $G_{ij}$, we get:
\begin{eqnarray}
\label{eq.evol} N \, g_{ij}(t+1)&=& N \, g_{ij}(t)+ \nonumber
\\&+& \sum_{k \neq j} g_{1j} \, g_{ik} \, \beta_{1i}+ \cdots +
\,\,
\sum_{k \neq j} g_{rj} \, g_{ik} \, \beta_{ri} \nonumber \\
&-&\sum_{k \neq j} g_{1k} \, g_{ij} \, \beta_{1i} - \cdots
-\sum_{k \neq j} g_{rk} \, g_{ij} \, \beta_{1i}, \nonumber
\\\,\,\, & & 0<i,j<r.
\end{eqnarray}

The explicit temporal dependence of the $g$'s has been left out in
order to avoid an overcharged notation. Each positive sum
represents the total probability that a transition \emph{to}
$G_{ij}$ is induced by a member of each authority group.
Conversely, each negative sum represents the probability that a
transition \emph{from} $G_{ij}$ is induced by a member of each
authority group. Grouping terms and using the conditions
$\sum_{j=1}^N g_{ij}=a_i$ we obtain:
\begin{eqnarray}
\label{eq.evol2} N \, g_{ij}(t+1)&=& N \, g_{ij}(t)+ \nonumber
\\& & -g_{ij}(t) \, \sum_{k \neq i} a_k \, \beta_{ki} +a_i \,
\sum_{k \neq j} g_{kj}(t) \, \beta_{ki}, \nonumber \\
\nonumber \\\,\,\, & & 0<i,j<r.
\end{eqnarray}
from where Eq.~(\ref{eq:mf}) follows.


\begin{thebibliography}{99}
\bibitem{weisbuch02} G. Weisbuch, G. Deffuant, F. Amblard, J. P. Nadal,
Complexity {\bf 7}, No. 2, 55 (2002).

\bibitem{stone61} M. Stone, Ann. of Math. Stat. {\bf 32}, 1339 (1961).

\bibitem{hegselmann02} R. Hegselmann and U. Krause, Journal of
Artificial Societies and Social Simulation \textbf{5}, no. 3, 2
(2002).

\bibitem{laguna03} M. F. Laguna, G. Abramson and D. H. Zanette,
Physica A \textbf{329}, 459 (2003).

\bibitem{laguna04} M. F. Laguna, G. Abramson and D. H. Zanette,
Complexity (in the press, 2004).

\bibitem{deffuant02} {G. Deffuant, F. Amblard, G. Weisbuch and T.
Faure, Journal of Artificial Societies and Social Simulation
\textbf{5}, no. 4, 1 (2002).}

\bibitem{bonabeau95} E. Bonabeau, G. Theraulaz and J.L. Deneubourg, Physica A
\textbf{217}, 373 (1995).

\bibitem{stauffer03} D. Stauffer and J.S. S\'{a} Martins, arXiv:cond-mat/0308437
(2003).

\bibitem{boguna03} M. Bogu\~{n}\'{a}, R. Pastor-Satorras, A.
D\'{\i}az-Guilera, and A. Arenas, arXiv:cond-mat/0309263 (2003).

\bibitem{copelli02} M. Copelli, R. M. Zorzenon dos Santos, J. S. Sa
Martins, Int. J. Mod. Phys. C \textbf{13}, 783 (2002).

\bibitem{tilman94} D. Tilman, R. M. May, C. L. Lehman and M. A.
Nowak, Nature \textbf{371}, 65 (1994).

\bibitem{note} According to the Frobenius-Perron theorem
(as applied to the matrix $(\mathbf{I}+\mathbf{A}/N)$ , this will
be the only eigenvector corresponding to the null eigenvalue,
provided that $\mathbf{A}$ is such that opinions are allowed to
``flow'' to every authority level. This is equivalent to demanding
that the evolution cannot be decoupled, which is the case of the
evolutions analyzed in this work.

\end{thebibliography}
\end{document}